\documentclass{emulateapj}  

\usepackage{graphicx} 
\slugcomment{Accepted in ApJ Letters}
\shorttitle{Quiet sun magnetic fields from space-borne observations} 
\shortauthors{Orozco Su\'arez et al.\/}

\begin{document}

\title{Quiet Sun magnetic fields from space-borne observations: simulating 
Hinode's case} 
\author{D.\ Orozco Su\'arez, L.R.\ Bellot Rubio, and J.C.\ del Toro Iniesta}

\affil{Instituto de Astrof\'{\i}sica de Andaluc\'{\i}a (CSIC), Apdo.\ 3004, 
18080 Granada, Spain; orozco@iaa.es, lbellot@iaa.es, jti@iaa.es}

\begin{abstract} 
We examine whether or not it is possible to derive the field strength
distribution of quiet Sun internetwork regions from very high spatial
resolution polarimetric observations in the visible.  In particular,
we consider the case of the spectropolarimeter attached to the Solar
Optical Telescope aboard Hinode.  Radiative magneto-convection
simulations are used to synthesize the four Stokes profiles of the
\ion{Fe}{1} 630.2~nm lines. Once the profiles are degraded to a
spatial resolution of 0\farcs32 and added noise, we infer the
atmospheric parameters by means of Milne-Eddington inversions. The
comparison of the derived values with the real ones indicates that the
visible lines yield correct internetwork field strengths and magnetic
fluxes, with uncertainties smaller than $\sim$150~G, when a stray
light contamination factor is included in the inversion.  Contrary to
the results of ground-based observations at 1\arcsec\/, weak fields
are retrieved wherever the field is weak in the simulation.
\end{abstract}


  \section{Introduction}
  \label{sec:intro}
The characterization of quiet sun internetwork (IN) fields is an important
issue in solar physics. Polarimetric measurements of the visible \ion{Fe}{1}
lines at 630.2~nm and the near-infrared \ion{Fe}{1} lines at 1565~nm have been
used to advance our knowledge of IN magnetism, but no consensus has been
reached yet.  While the analysis of the visible lines suggest a predominance
of kG field strengths and small filling factors (S\'anchez Almeida \& Lites
2000; Dom\'{\i}nguez Cerde\~na et al.\ 2003; Socas-Navarro \& Lites 2004), the
near-infrared lines indicate that most fields have hG strengths with
larger filling factors (Lin 1995; Lin \& Rimele 1999; Khomenko et al.\ 2003;
Mart\'{\i}nez Gonz\'alez et al.\ 2006; Dom\'{\i}nguez Cerde\~na et al.\ 2006).

Attempts to reconcile these contradictory results have argued that visible and
IR lines sample different magnetic structures in the resolution element
(S\'anchez Almeida \& Lites 2000; Socas-Navarro \& S\'anchez Almeida 2003) or
that noise affects the visible lines more dramatically than the IR lines
(Be\-llot Rubio \& Collados 2003). On the other hand, Mart\'{\i}nez Gonz\'alez
et al.\ (2006) have convincingly demonstrated that it is not possible to
obtain reliable IN field strengths from the \ion{Fe}{1} 630.2~nm lines due to
crosstalk with thermodynamical parameters at 1\arcsec\/ resolution.

Different techniques have been proposed to study IN fields, but the diagnostic
potential of high spatial resolution observations in the absence of
atmospheric seeing has only been explored by Khomenko et al.\ (2004).
The spectropolarimeter (SP; Lites et al.\ 2001) of the Solar Optical Telescope
(SOT) aboard Hinode (Ichimoto et al.\ 2005) is already providing nearly
diffraction-limited observations of the solar photosphere, and upcoming
instruments will do so in the future (e.g., IMaX aboard SUNRISE or VIM 
aboard Solar Orbiter). Thus, there is a clear need to assess whether 
reliable IN field strengths can be derived from space-borne observations.

Here we address this question by simulating and analyzing
Hinode/SP measurements. Radiative magneto-convection simulations are
used to synthesize the Stokes profiles of the \ion{Fe}{1}
630.2~nm lines. The profiles are degraded to the nearly
diffraction-limited resolution of 0\farcs32 achieved by
Hinode/SP. After adding noise to the Stokes spectra, the atmospheric
parameters are inferred by means of inversion techniques. By comparing
the derived values with the real ones we determine the errors in field
strength and magnetic flux to be expected from the analysis of Hinode
data. Our main result is that Milne-Eddington (ME) inversions of the
visible 630.2~nm lines yield correct IN field strengths
with uncertainties smaller than 150~G for the whole range of strengths
from $\sim$0.1 to 1~kG.  If internetwork fields are hG fields, then
simple ME inversions of Hinode/SP measurements will result in sub-kG
field strength distributions, contrary to what is obtained from
current ground-based observations.

%

\section{MHD simulations, spectral synthesis, and image degradation}
\label{sec:simul}

To describe the sun in the more realistic way possible we use
radiative MHD simulations by V\"ogler et al.\ (2005).  Specifically,
we take three snapshots from different simulation runs representing
very quiet, unipolar internetwork and network regions with average
unsigned fluxes of 10, 50, and 200~Mx~cm$^{-2}$, respectively. The
horizontal and vertical extents of the computational box are 6 and
1.4~Mm. The synthesis of the Stokes spectra of the two \ion{Fe}{1}
lines is carried out using the SIR code (Ruiz Cobo \& del Toro Iniesta
1992). The spectral region is sampled at 113 wavelength positions in
steps of 2.15~pm, following the Hinode/SP normal map mode (for
details, see Shimizu 2004). The atomic parameters have been taken from
the VALD database (Piskunov et al.\ 1995).

\begin{figure*}
\centering 
\epsscale{0.38} 
\plotone{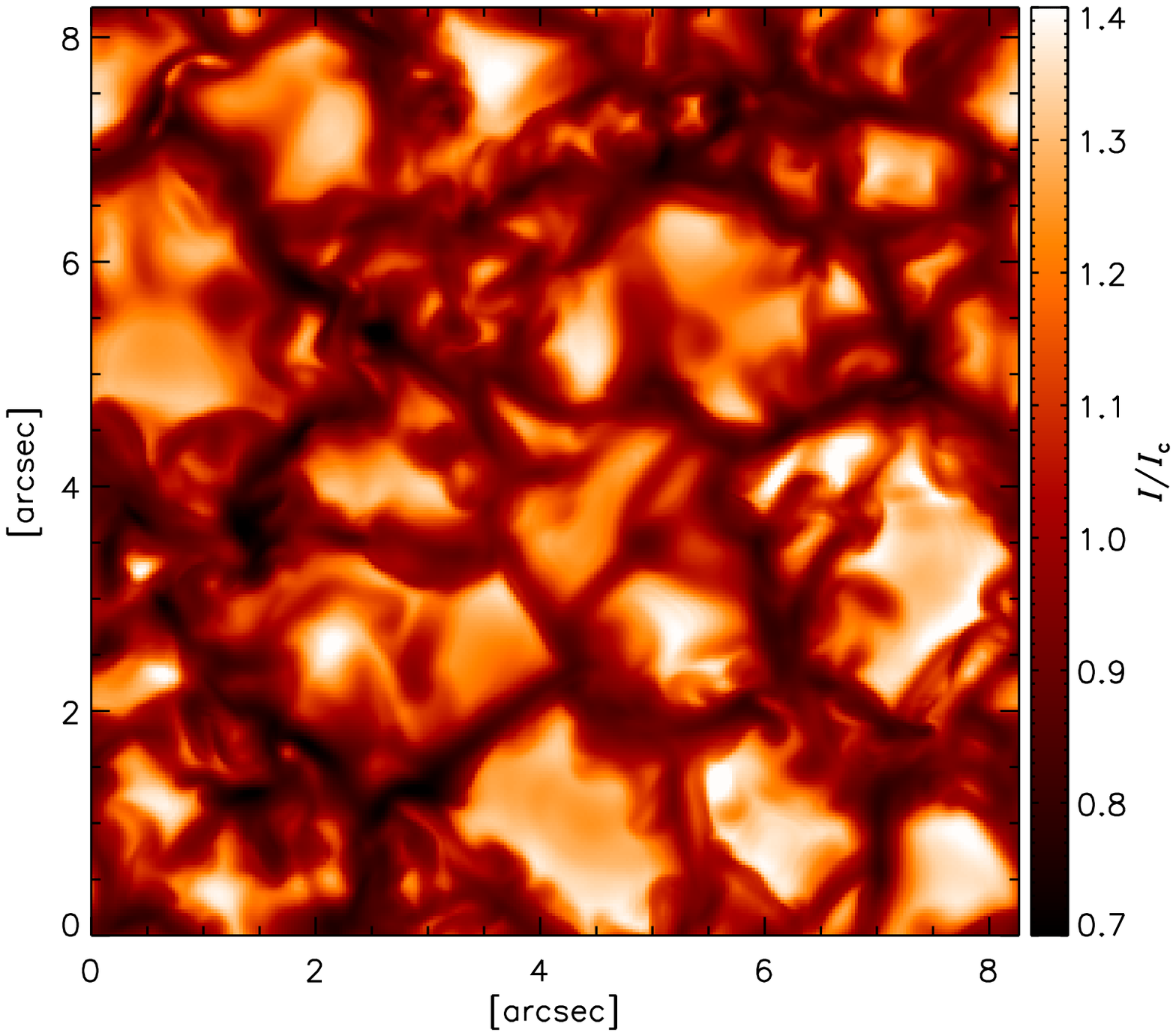} 
\plotone{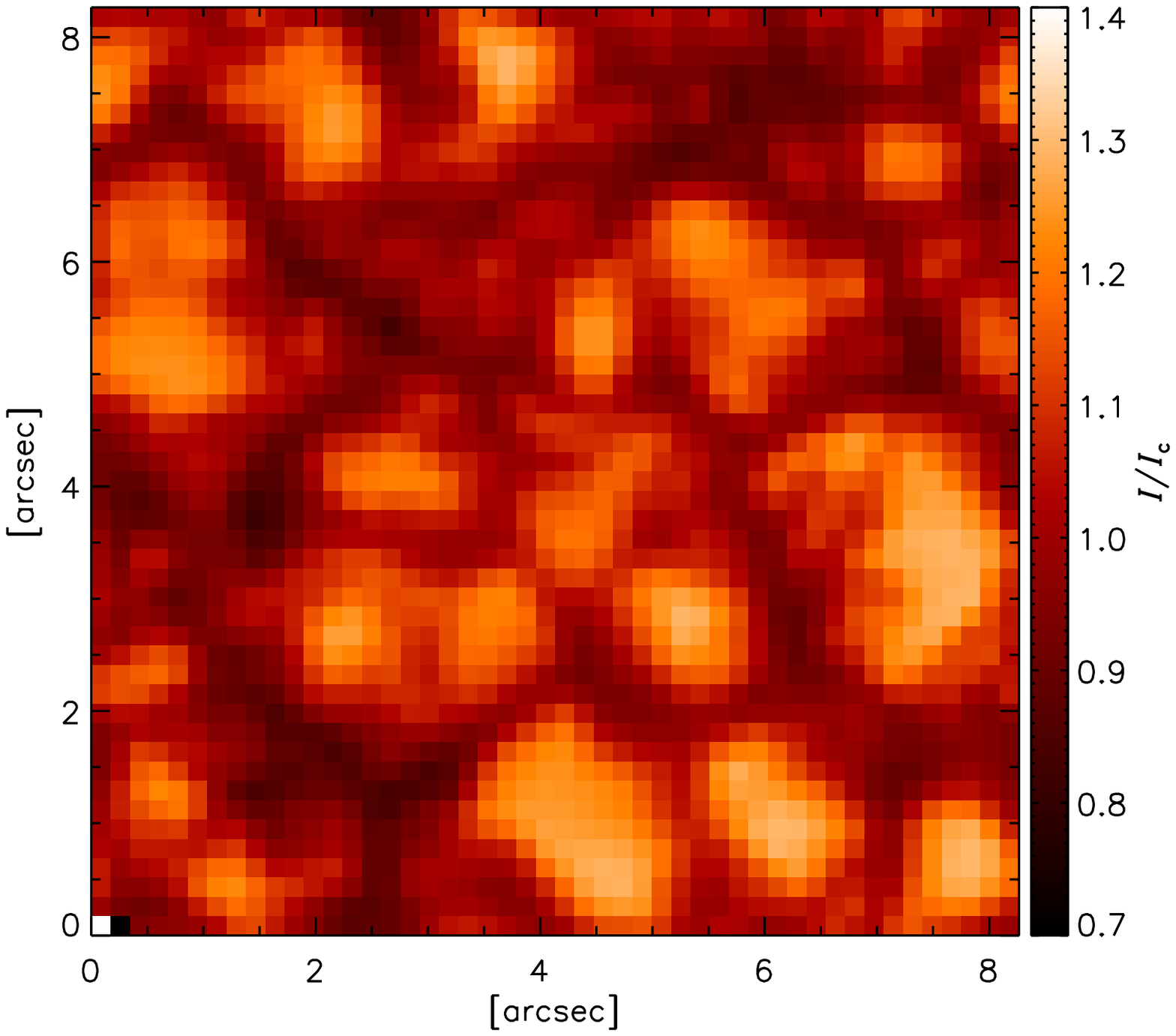}
\plotone{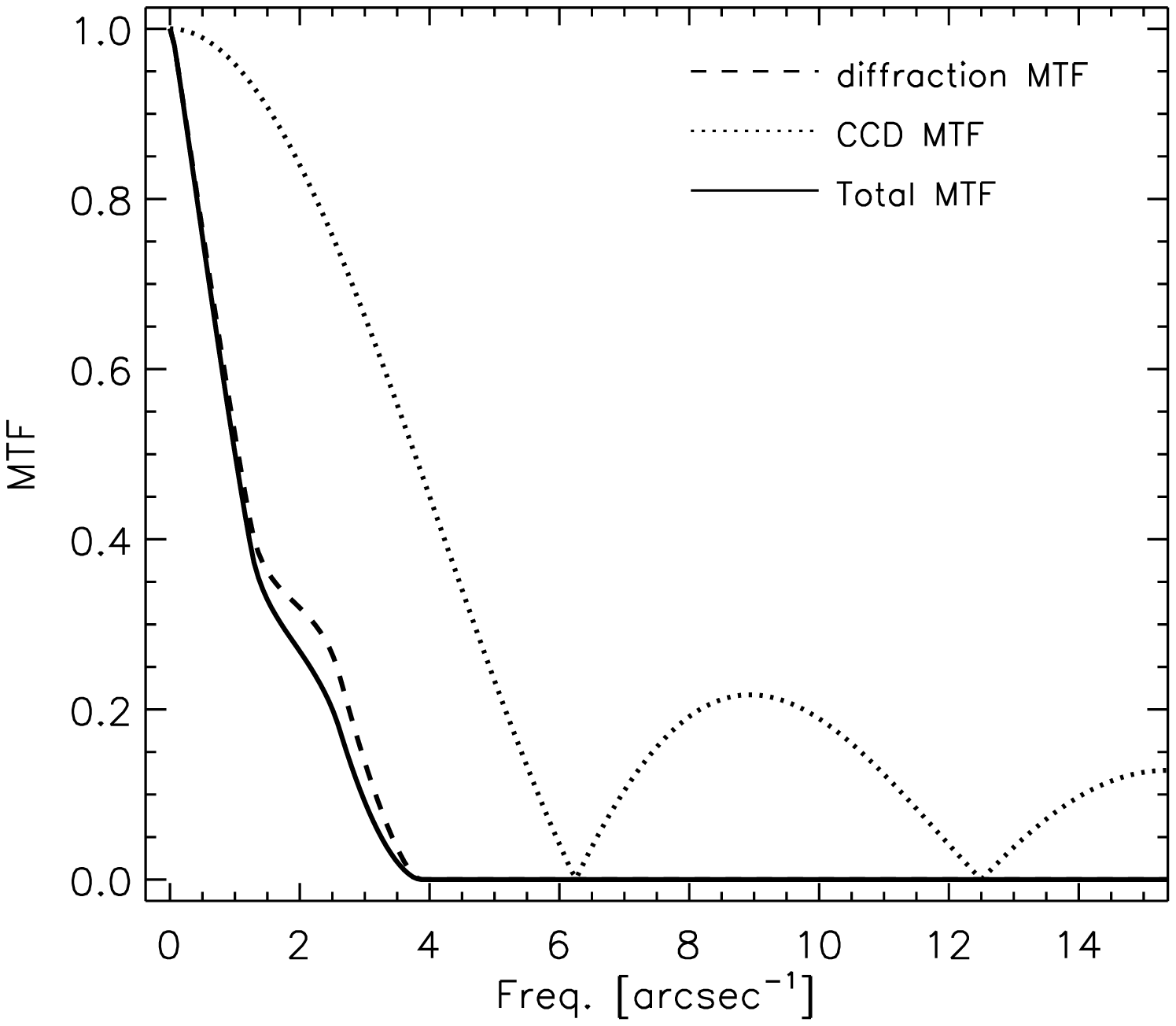}
\caption{{\em Left and middle:} Continuum intensity maps for the
simulation snapshot with average unsigned flux of 10~Mx~cm$^{-2}$ 
and for the data spatially degraded considering telescope diffraction 
and pixel size. Color scales are the same in the two maps. The contrast 
varies from 13.7\% in the original image to 8.5\% in the degraded one. 
{\em Right:} MTF of the detector (dotted line), diffraction limited MTF 
(dashed line), and combination of both effects (solid line).}
\label{fig:fig1}
\end{figure*}

The aperture of Hinode/SOT is 0.5~m, which operating at 630~nm provides a
spatial resolution of $\sim$0\farcs26 (equivalent to $\sim$\,190~km on the
solar surface). The sampling interval in the MHD simulations is 0\farcs0287,
implying a spatial resolution of 0\farcs057 (41.6~km). Thus, in order to
simulate SOT observations, the synthetic Stokes profiles derived from the MHD
model have been spatially degraded by telescope diffraction. In addition, we
have considered the extra loss of contrast caused by the integration of the
signal in the detector. Finally, the images are rebinned to the SP CCD pixel
size of 0$\farcs$16$\times$ 0$\farcs$16.  Figure~\ref{fig:fig1} shows
continuum intensity maps for the original and the spatially degraded data. 
The degradation process reduces the rms contrast from $\sim$14\% to $\sim$9\% 
in the continuum. The pixelation of the CCD is noticeable in the degraded
image. The right panel of Fig.~\ref{fig:fig1} shows Modulation Transfer
Functions (MTFs) describing the filtering of spectral components induced 
by telescope diffraction and pixelation effects in the CCD.  Note the
modification of the effective MTF caused by the central obscuration of the
entrance pupil.

\begin{figure}
\centering
\epsscale{0.8}
\plotone{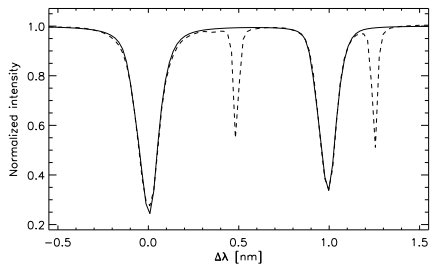}
\caption{Comparison between the average Stokes $I$ profile from the
spatially degraded data (solid) and the FTS spectral atlas
(dashed). Both continua are normalized to unity.}
\label{fig:fig2}
\end{figure}

Figure~\ref{fig:fig2} compares the average Stokes $I$ profiles from the
spatially degraded data and the NSO Fourier Transform Spectrometer Atlas of
the quiet Sun. Both spectra are very similar, with only small differences in
the line core and wings of \ion{Fe}{1} 630.1~nm. Note that the lack of a
temporal average exclude, for instance, the effect of the 5-min oscillation 
in the simulated profile. To reproduce an Hinode/SP observation we have also
convolved the profiles with a Gaussian of 30 m{\AA}~FWHM to account for the
spectral resolving power of the spectrograph and have added noise at
the level of 10$^{-3}$ of the continuum intensity $I_{\rm c}$ 
(the polarimetric sensitivity of standard Hinode/SP measurements).

%

\section{Inversion}

To derive the magnetic field strength from the simulated profiles we
use a least-square inversion technique based on ME atmospheres.  ME
inversions represent the best option to interpret the measurements if
one is not interested in vertical gradients of the physical
quantities.  They are simple and often provide reasonable averages of
the atmospheric parameters over the line formation region (Westendorp
Plaza et al.\ 2001; Bellot Rubio 2006).

We apply the ME inversion to the \ion{Fe}{1} 630.15~nm and \ion{Fe}{1} 
630.25~nm lines simultaneously. A total of 9 free parameters are 
determined (S$_0$, S$_1$, $\eta_0$, $\Delta\lambda_D$, $a$, $B$, 
$\gamma$, $\chi$, and v$_\mathrm{LOS}$; for the meaning of the symbols 
see, e.g., Orozco Su\'arez \& del Toro Iniesta 2007). 
No additional broadening of the profiles by macroturbulence 
or microturbulence is allowed. 
Three different inversions are performed to derive the atmospheric
parameters. All of them use a simple one-component model, i.e., a
laterally homogeneous magnetic atmosphere occupying the whole resolution
element. We first invert the profiles in the absence of noise, and
then with noise added at the level of 10$^{-3} \, I_{\rm c}$. In the
last inversion, the noisy profiles are fitted considering non-zero
stray light contaminations factors.
The stray light profile is evaluated individually for each pixel by 
averaging the Stokes $I$ profiles within a box 1\arcsec\/-wide centered 
on the pixel. For all inversions we use the same initial guess model, 
allowing a maximum of 300 iterations. The initial field strength is 100~G. 

%

\section{Results}
\label{sec:res}

\begin{figure*}
\centering
\epsscale{1.1}
\plotone{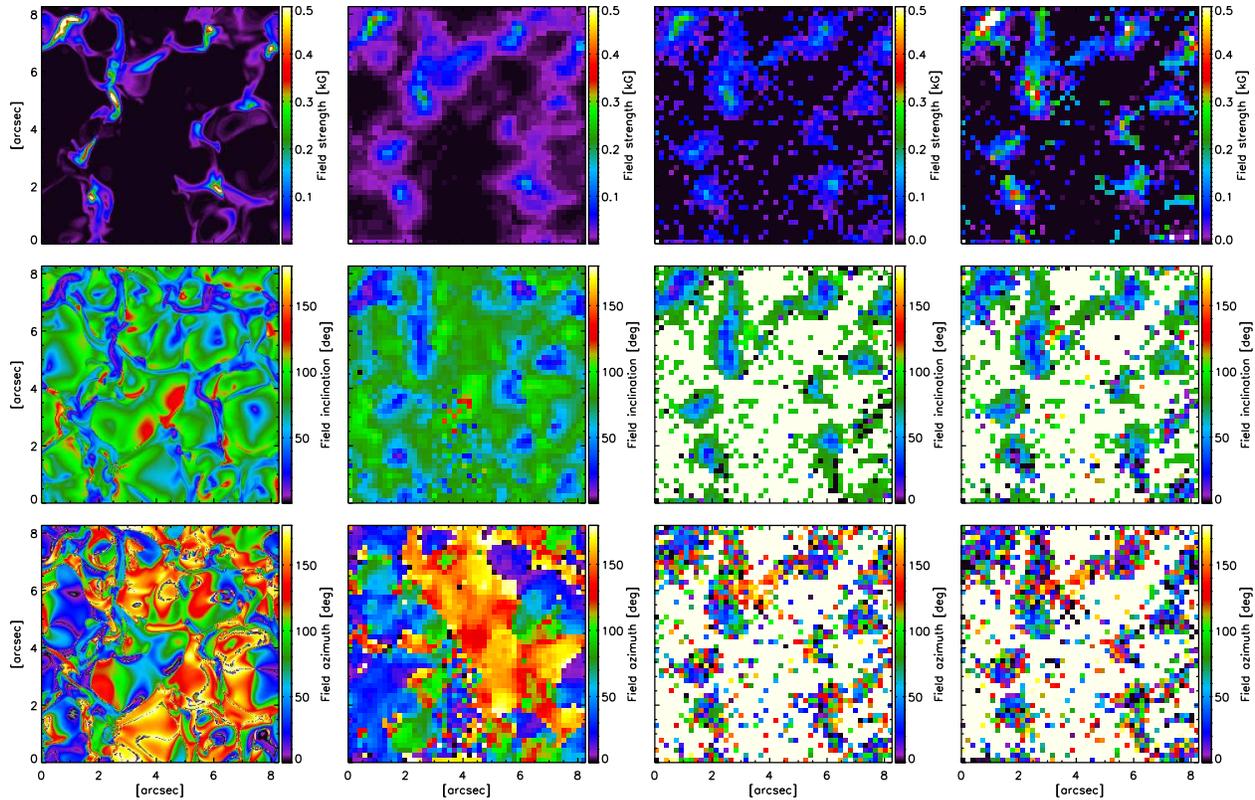}
\vspace{-2em}
\caption{{\em Left}: Cuts at optical depth ${\rm log} \, \tau =-2$ of
the model atmospheres provided by the MHD simulation with averaged
unsigned flux of 10~Mx~cm$^{-2}$. {\em Second column}: Maps of the
physical quantities retrieved from the ME inversion of the profiles
with no noise. {\em Third column}: Maps retrieved from the ME
inversion of the profiles with SNR of 1000 and no stray light
contamination. {\em Fourth column:} Same as before, but accounting for
stray light. From top to bottom: magnetic field strength, inclination,
and azimuth.}
\label{fig:mapas}
\end{figure*}

Figure~\ref{fig:mapas} shows the vector magnetic field (strength,
inclination, and azimuth) retrieved from the inversions of the Stokes
profiles. The first column displays a cut of the simulation snapshot
with average flux density of 10~Mx~cm$^{-2}$ at optical depth ${\rm
log} \, \tau=-2$.  The second and third columns contain the results of
the ME inversions of the spatially degraded profiles in the absence of
noise and the specific case of a SNR of 1000, respectively. Finally,
the fourth column shows the atmospheric parameters derived from the
noisy profiles accounting for stray light contamination. White regions
represent pixels which have not been inverted because of their small
polarization signals (we only consider pixels whose Stokes $Q$, $U$ or
$V$ amplitudes exceed three times the noise level).

Over the granules, the magnetic field strength is very weak and the
polarization signals are buried in the noise. These pixels represent
$\sim$55\% of the total area (white regions in Fig.~\ref{fig:mapas}).
The stronger fields concentrate in intergranular regions. In those
regions, the magnetic structures inferred from the inversion have
bigger sizes than the real ones, i.e., they appear ''blurred''. This 
is caused by the degradation of the images due to telescope 
diffraction and CCD pixel size.  The field inclination and 
azimuth structures resulting from the inversion are blurred as 
well. The azimuth values are rather uncertain because of the tiny 
linear polarization signals produced by the weak fields of the 
simulations.

\begin{figure}[!t]
\centering
\epsscale{1.1}
\plotone{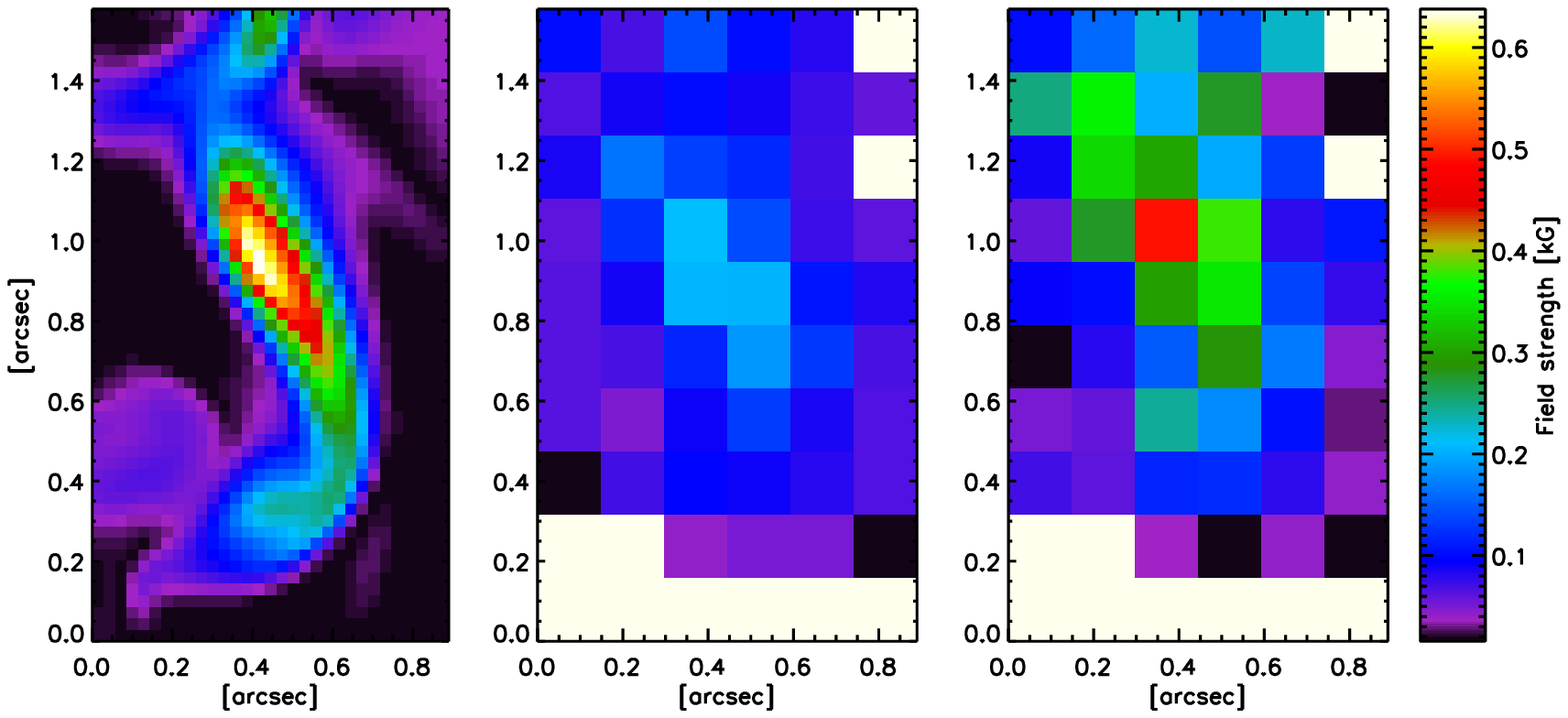}
\plotone{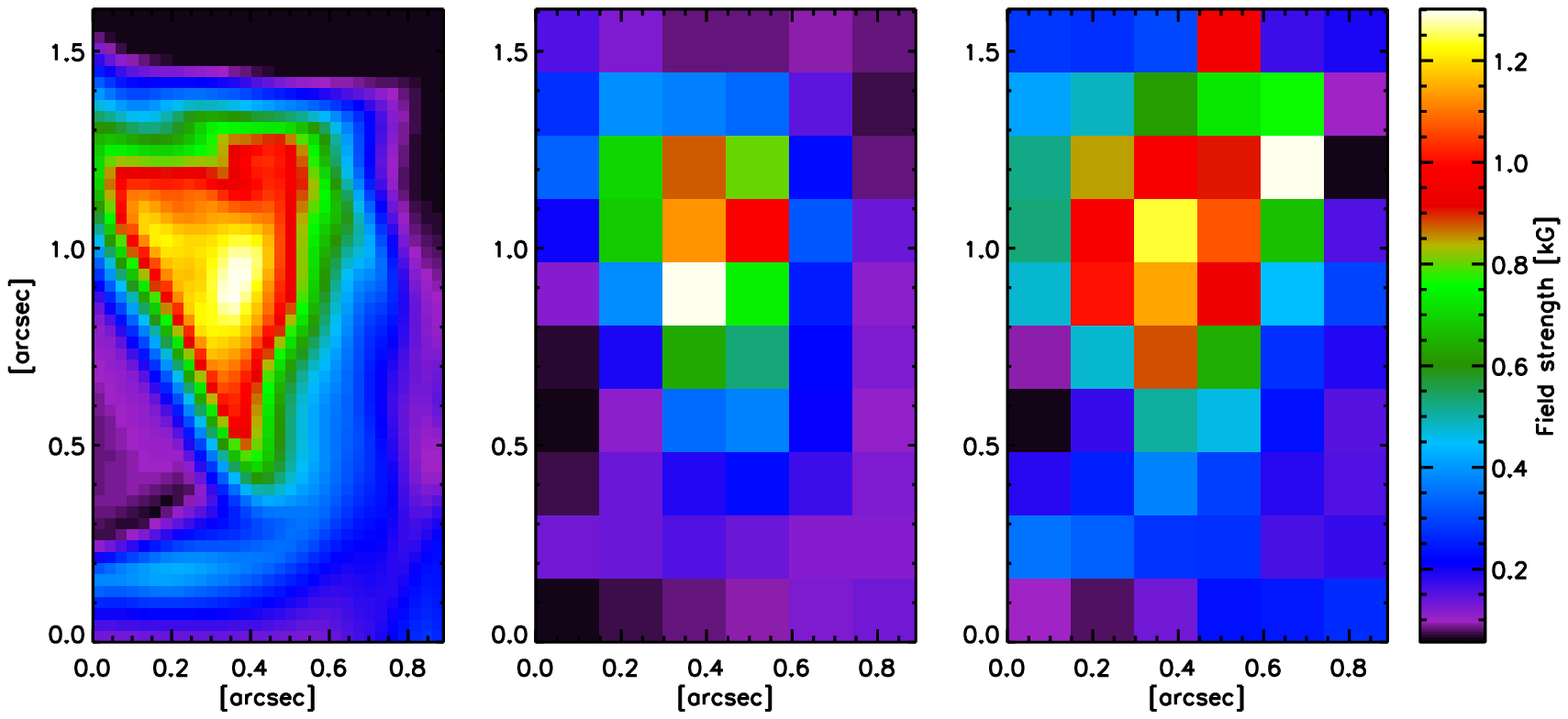}
\caption{{\em Left:} Field strengths at ${\rm log} \, \tau =-2$ in the
MHD simulations with 10~Mx~cm$^{-2}$ (top) and 50~Mx~cm$^{-2}$
(bottom). {\em Middle}: Field strengths derived from the ME inversion
of the spatially degraded Stokes profiles with SNR 1000 and no stray
light contamination. {\em Right}: Field strengths from the ME
inversion accounting for stray light contamination. }
\label{fig:mapaszoom}
\end{figure}

Figure \ref{fig:mapaszoom} is a close up of small features observed in
intergranular regions. Note that each Hinode/SP pixel of $0\farcs16
\times 0\farcs16$ corresponds to 36 pixels in the simulation, hence
they usually contain a broad distribution of magnetic field
strengths. When we consider that the polarization signal is produced
by a single magnetic component within the resolution element and no
stray light is allowed for, the inferred field strengths are smaller
than those in the model, so the field is underestimated (middle panels
of Fig.~\ref{fig:mapaszoom}). If one accounts for stray light
contamination the inferred fields become stronger (right panels), but
also noisier due to the increased number of free parameters.

\begin{figure}
\centering 
\epsscale{1.1}
\plotone{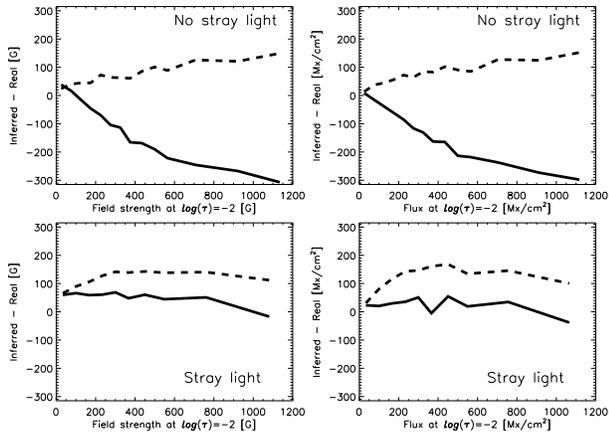}   
\vspace*{.5em}
\caption{{\em Top:} Mean (solid) and rms (dashed) errors of the field
strength (left) and flux (right) derived from the inversion of the
profiles with SNR=1000 assuming a single magnetic atmosphere and no
stray light contamination. {\em Bottom:} Same as before but accounting
for stray light contamination.}
\label{fig:fig5}
\end{figure}

To analyze the results in a more quantitative way we calculate the mean 
and rms values of the errors. We define the error as the difference
between the inferred and the real parameters at optical depth ${\rm
log} \, \tau=-2$. Since one pixel of the degraded data corresponds 
to 36 pixels in the simulations, we compare each inverted pixel with 
the mean of the corresponding 36 pixels in the original map. 
Figure~\ref{fig:fig5} shows the mean and rms errors of the field 
strength resulting from the inversion without accounting for stray
light (top left panel). It is clear that fields above $\sim$100~G are
underestimated, with rms errors smaller than $\sim$150~G in the whole
range of strengths. The results are similar for the magnetic flux
density (top right panel). The inversion considering stray light
contamination yields much better inferences, as can be seen in the
bottom panels of Fig.~\ref{fig:fig5}. The field strength and flux 
are slightly overestimated for weak fields, but the rms errors 
do not exceed 150~G in any case.

%

\section{Discussion}

Previous analyses of visible (630.2~nm) and near-infrared (1565~nm)
neutral iron lines do not agree on the distribution of field strengths
in IN regions. In particular, the visible lines systematically deliver
kG field strengths and small filling factors, while the near-infrared
lines suggest a predominance of hG fields.

Here we have shown that ME inversions of the \ion{Fe}{1}~630~nm lines at
spatial resolutions of 0\farcs32 (the case of Hinode/SP) underestimate the
magnetic field strength by some hundred G if no stray light contamination is
included.  When stray light is accounted for, ME inversions are able
to recover any magnetic field above 100~G with remarkable accuracy.

Interestingly, we always derive weak fields from the simulated
Hinode/SP observations where the field in the MHD model is
weak. Likewise, pixels assigned strong fields by the ME inversion
actually correspond to strong fields in the MHD model. This is in
sharp contrast with the results of Mart\'{\i}nez Gonz\'alez et al.\
(2006), who always infer kG fields from the \ion{Fe}{1} 630.2~nm 
lines observed in the IN at resolutions of 1\arcsec\--1\farcs5 when 
the inversion is initialized with strong fields. The difference is
probably due to: (a) our significantly higher spatial resolution,
which narrows the range of field strengths present in the pixel; (b)
the fact that we do not employ two-component atmospheres, micro- or
macro-turbulent velocities, which reduces the degrees of freedom of
the solution; and (c) the simple description of the thermodynamics
provided by the ME model which, contrary to the atmosphere used by
Mart\'{\i}nez Gonz\'alez et al.\ (2006), does not allow to compensate
for incorrect magnetic parameters.  On the other hand, our results
seem to contradict the conclusions of Bellot Rubio \& Collados
(2003). However, the signals considered here are larger by a factor 
of $\sim$10 due to the much higher angular resolution (which implies
larger filling factors). Under these conditions, noise does not
significantly affect the field strengths derived from the visible
lines (see Fig.\ 5 in Bellot Rubio \& Collados 2003). 

We caution that the results of this Letter may only be valid as 
long as the MHD simulations provide a realistic description of the Sun.  
The performance of ME inversions could be different if the magnetic field 
is structured on scales much smaller than $\sim$0\farcs3. For the moment,
however, there is no compelling evidence that tiny magnetic elements 
exist in the quiet solar photosphere.


  \section{Conclusions}
  \label{sec:con}

Our analysis suggests that Hinode/SP observations will make it possible to
determine the real distribution of field strengths in quiet Sun internetwork
regions. Simple one-component Milne-Eddington inversions without macro and
microturbulence seem appropriate to achieve that goal. However, it will be
essential to account for the degradation of the image induced by telescope
diffraction and detector pixel size. The work presented here shows that the
effects of the degradation can be modeled sufficiently well including a 
stray light contamination factor in the inversion.

%

\acknowledgments We thank A.\ V\"ogler and M.\ Sch\"ussler for making
their MHD simulations available and answering our questions about
them. The program used to degrade the synthetic Stokes profiles was
written by J.A.\ Bonet and S.\ Vargas. This work has been partially
funded by the Spanish Mi\-nisterio de Educaci\'on y Ciencia through
project ESP2003-07735-C04-03 (in\-clu\-ding European FEDER funds) and
{\em Programa Ram\'on y Cajal}.

%

%


\begin{thebibliography}{}



\bibitem[Bellot Rubio(2006)]{spw4} Bellot Rubio, L.~R.\
2006, ASP Conf.\ Ser.: Solar Polarization 4, 358, 107

\bibitem[Bellot Rubio \& Collados(2003)]{2003A&A...406..357B} Bellot Rubio, 
L.~R., \& Collados, M.\ 2003, \aap, 406, 357 

\bibitem[]{} 
Dom{\'{\i}}nguez Cerde{\~n}a, I., Kneer, F., \& S{\'a}nchez Almeida, J.\ 
2003, \apjl, 582, L55  

\bibitem[Dom{\'{\i}}nguez Cerde{\~n}a et al.(2006)]{2006ApJ...646.1421D} 
Dom{\'{\i}}nguez Cerde{\~n}a, I., Almeida, J.~S., \& Kneer, F.\ 2006, \apj, 
646, 1421 


\bibitem[Ichimoto \& Solar-B Team(2005)]{2005JKAS...38..307I} Ichimoto, K., 
\& Solar-B Team 2005, Journal of Korean Astronomical Society, 38, 307 

\bibitem[Khomenko et al.(2003)]{2003A&A...408.1115K} Khomenko, E.~V., 
Collados, M., Solanki, S.~K., Lagg, A., \& Trujillo Bueno, J.\ 2003, \aap, 
408, 1115 

\bibitem{} Khomenko, E., Collados, M., \& Solanki, S.K.\ 2004, Mem.\ S.A. It., 
75, 282

\bibitem[Lin(1995)]{1995ApJ...446..421L} Lin, H.\ 1995, \apj, 446, 421 

\bibitem[Lin \& Rimmele(1999)]{1999ApJ...514..448L} Lin, H., \& Rimmele, 
T.\ 1999, \apj, 514, 448 

\bibitem[Lites et al.(2001)]{2001ASPC..236...33L} Lites, B.~W., Elmore, 
D.~F., \& Streander, K.~V.\ 2001, ASP Conf.~Ser.: Advanced Solar 
Polarimetry -- Theory, Observation, and Instrumentation, 236, 33 


\bibitem[Mart{\'{\i}}nez Gonz{\'a}lez et al.(2006)]{2006A&A...456.1159M} 
Mart{\'{\i}}nez Gonz{\'a}lez, M.J., Collados, M., \& Ruiz Cobo, B.\ 2006, 
\aap, 456, 1159 

\bibitem[Orozco Su{\'a}rez \& del Toro Iniesta(2007)]{2007A&A...462.1137O} 
Orozco Su{\'a}rez, D., \& del Toro Iniesta, J.C.\ 2007, \aap, 462, 1137 

\bibitem[Piskunov et al.(1995)]{piskunov} Piskunov, N.~E.,
Kupka, F., Ryabchikova, T.~A., Weiss, W.~W., \& Jeffery, C.~S.\ 1995,
A\&A Supp.\ Ser., 112, 525

\bibitem[Ruiz Cobo \& del Toro Iniesta(1992)]{1992ApJ...398..375R} Ruiz 
Cobo, B., \& del Toro Iniesta, J.~C.\ 1992, \apj, 398, 375 


\bibitem[]{} S\'anchez Almeida, J., \& Lites, B.W.\ 2000, \apj, 532, 1215 

\bibitem[Shimizu(2004)]{2004ASPC..325....3S} Shimizu, T.\ 2004, ASP 
Conf.~Ser.: The Solar-B Mission and the Forefront of Solar Physics, 
325, 3 

\bibitem[]{}
Socas-Navarro, H., \& S\'anchez Almeida, J.\ 2003, \apj, 593, 581 

\bibitem[]{} Socas-Navarro, 
H., \& Lites, B.~W.\ 2004, \apj, 616, 587 


\bibitem[V{\"o}gler et al.(2005)]{voegler} V{\"o}gler, A.,
Shelyag, S., Sch{\"u}ssler, M., Cattaneo, F., Emonet, T., \& Linde,
T.\ 2005, A\&A, 429, 335

\bibitem[Westendorp Plaza et al.(2001)]{carlos} Westendorp
Plaza, C., del Toro Iniesta, J.C., Ruiz Cobo, B., Mart\'{\i}nez Pillet, V., 
Lites, B.W., \& Skumanich, A.\ 2001, ApJ, 547, 1130


\end{thebibliography}
\end{document}